\documentstyle[prl,aps]{revtex}

\input epsf
\tighten

\begin{document}
\title{Non-topological solitons in brane world models
}
\author{Dejan Stojkovic}
\address{Department of Physics, University of Alberta, Edmonton, AB T6G
2J1, Canada \\
e-mail: dejans@ualberta.ca}

\vspace{.8cm}

\wideabs{
\maketitle

\begin{abstract}
\widetext We examine some general properties of a certain class of
scalar filed theory models containing non-topological soliton
solutions in the context of brane world models with compact large
extra dimensions. If a scalar field is allowed to propagate in
extra space, then, beside standard Kaluza-Klein type excitations,
a whole new class of very  massive soliton-type states can exist.
Depending on their abundance, they can be important dark matter
candidates or give significant contribution to entropy and energy density in
our universe.

\end{abstract}
}

\narrowtext

It is well known  that non-linear scalar filed theory models have
a very rich structure. The scalar filed theory, with only linear
couplings, admits only plane wave solutions. Wave packets can be
constructed by superposition of plane waves but they are always
dispersive. If non-linear couplings are present, then the theory
can contain spatially confined nondispersive solutions ---
solitons. In renormalizable relativistic local field theories,
all the  soliton solutions belong to two groups --- topological
and non-topological solitons \cite{Lee}.

The necessary condition for the existence of a topological soliton
is that there should exist degenerate vacuum states so that the
boundary conditions at infinity for a soliton state are
topologically different from that of a physical vacuum state.
There are well known examples of such solutions in field theory
--- monopoles, strings, domain walls etc. Their stability is
insured against decay into plain waves because of topologically
non-trivial boundary conditions at infinity.

The necessary condition for the existence of a non-topological
soliton is that there should exist an additive conservation law.
The boundary conditions at infinity are the same as for the
vacuum state. Because of the absence of non-trivial topology, the
stability of non-topological solitons depends on which type of
solution is of the lowest energy\footnote{It is also possible to achieve
the stability by including some additional fields. For example, the stability of
electroweak strings can be provided by  fermionic zero modes in theory \cite{ews}}.

There is a simple result called Derrick's theorem \cite{Derrick}
that imposes severe restrictions on the number of scalar field
models which can contain solitons. In principle, if the number of
space-like dimensions in theory is greater than one, then the
only time independent solutions of finite energy are the vacuum
states which are constant everywhere and which minimize the
potential energy. There are several ways to go around this
result. In order to have soliton solutions, we can (1) include
fields with non-zero spin  (2) consider scalar models with
constraints (3) consider time-dependent but non-dispersive
solutions.

In literature, a great part of attention was devoted to
topological solitons in arbitrary number of space-time dimensions
with both compact and non-compact extra dimensions. However,
analysis of non-topological soliton solutions in literature was
mainly limited  to  calculations based on assumption that the
space-time is $(3+1)$-dimensional on all scales and that the
strength of the gravitational interaction is given by the Plank
mass scale, $M_{Pl}$ \cite{Lee,ss}. The goal of this paper is to
present some general features of non-topological solitons in the
context of recently proposed theories with large compact extra
dimensions. We will show that new assumptions about the
dimensionality of our space-time and the gravitational strength
at small distances allows the existence of two new types of
particles which can possibly play a significant role in cosmology.

Since we are mainly interested in questions of importance for
cosmology, we analyze non-linear scalar field theory models in the
presence of gravity. Consider gravitational Lagrangian in $(4+d)$
dimensions with a complex massive scalar field $\Phi$:

\begin{equation}  \label{L}
L = \int{\ d^{4+d} x \sqrt{-g^{(4+d)}} \left[ {\cal R}^{(4+d)} + g^{\mu \nu}
\partial_\mu \Phi \partial_\nu \Phi + U( \Phi ) \right] }
\end{equation}
where ${\cal R}$ is a Ricci scalar and $U(\Phi)$ is a potential
for the field $\Phi$. The field $\Phi$ propagates in all $(4+d)$
dimensions. There are two distinct classes of non-topological
soliton solutions in these models. The necessary condition for
both of them is that the theory is invariant under space
independent phase rotations:

\begin{equation}  \label{u1}
\Phi \rightarrow \Phi^{\prime}= e^{i \alpha } \Phi
\end{equation}
This provides conservation of a quantity $N$ which we identify
with the particle number.  Class I comprise models where
$U(\Phi)$ is such that the theory does not contain
non-topological soliton solutions in the absence of the
gravitational field. Class II comprise models  where $U(\Phi)$ is
such that the theory does contain non-topological soliton
solutions in the absence of the gravitational field \cite{ss}.

Analysis of these models has been done only for the case of
$(3+1)$ space-time dimensions\footnote{After this paper
appeared on the web we noticed the complementary work in \cite{FST}.}. Here,
we examine models of class I and II in the context of recently proposed "world as a brane
scenario". We consider an original \cite{ADD} of large extra
dimensions (for other types of compactification of extra
dimensions see \cite{other}).
In this proposal, our universe is a direct product of an ordinary $(3+1)$%
-dimensional space and an extra compact $d$-dimensional space. All the
standard model fields are trapped on the brane which represents an ordinary $%
(3+1)$-dimensional space, while gravity can propagate everywhere.
This scenario allows a solution of so-called hierarchy problem by
bringing down a fundamental quantum gravity energy scale ---
$M_F$ --- down to the electroweak energy scale. For example, in
the case of toroidal compactification of the extra space, $M_F$
can be as low as $1$TeV if the radius of compactification is $r_c
\sim 1$mm in $d=2$. For an observer on the brane, the standard
Newton's law of gravity  remains valid on distances larger than
$r_c$, i.e. for $r >r_c$ the gravitational force, $F_g$, between
two bodies of mass $m_1$ and $m_2$ is governed by the
$4$-dimensional gravitational constant, $G_4 = 1/M_{Pl}^2$

\begin{equation}  \label{NL}
F_g = \frac{G_4 m_1 m_2}{r^2} \, .
\end{equation}

However, for $r < r_c$ the space is effectively $(4+d)$ dimensional and the
Newton's law is modified:

\begin{equation}
F_{g}=\frac{G_{4+d}m_{1}m_{2}}{r^{2+d}}  \label{MNL} \, ,
\end{equation}
where the $(4+d)$-dimensional gravitational constant
$G_{4+d}=1/M_{F}^{2+d}$. It is important to note that the
strength of gravitational interaction is now determined by a
fundamental energy scale, $M_{F}$, and not by the Plank scale,
$M_{Pl}$. We will see that this fact will considerably modify the
basic properties of soliton states.

For comparison, we give the basic features of $(3+1)$ and
$(3+1+d)$-dimensional models in parallel. The model which gives
rise to class I soliton solutions in $(3+1)$ dimensions is
described by the Lagrangian (\ref{L}) with $d=0$ and the
potential given by

\begin{equation}
U(\Phi) = m^2 \Phi^\dagger \Phi \,  ,
\end{equation}
where $m$ is the mass of the field $\Phi$ \cite{ss}. This model
supports a classical soliton solution which is regular everywhere
and zero at infinity. Its amplitude is determined by an
attractive force coupling constant. In this case the amplitude is
proportional to $1/\sqrt{G_{4}}$ since the only attractive force
in the model is gravity. Ignoring gravity loops, a renormalizable
quantum theory can be obtained in analogy with well established
procedures for quantum solitons.

Because of Bose-Einstein statistics, it is possible to have very
massive objects with the extremely compact characteristic
dimensions. The equations of motion can be derived by minimizing
the total (gravitational plus scalar field) energy while keeping
the number of particles, $N$, fixed \cite{ss}. The spherically
symmetric solution

\begin{equation}
\Phi = 1/\sqrt{2} \sigma (r) e^{-i \omega t}
\end{equation}
is regular everywhere for $\omega < m$ (non-topological soliton
can be considered as an analytical continuation of a plane wave
solution for which $\omega \geq m$ always)  and at infinity goes
to zero as

\begin{equation}
\sigma (r) \sim e^{-\sqrt{m^2-\omega^2} r} \rightarrow 0 \ .
\end{equation}

Thus, the characteristic size of a class I soliton object, in the
lowest energy state, is proportional to $m^{-1}$. Here, we assume
that $\Phi$ is some Higgs-type scalar field, and thus for the
usual range of the scalar field masses the linear size of a class
I soliton  is microscopic. It can be shown that, because of
gravitational interaction, for large $N$, these states are stable.

It is important to note that very reliable order of magnitude
estimates describing almost all of the peculiar properties of
non-topological solitons are possible to get without solving any
differential equations. To see this, examine the set of $N$
particles (solitons) in the same orbit in a lowest energy state (a
zero-node or s-state) of wavelength $R$. Define the effective
mass of this configuration by $M \sim N m$,where $m$ is the mass
of the free particle. The kinetic energy $\sim N/R$ prevents the
 collapse of the object, while gravitational energy $\sim G_4 M^2/R$
prevents its decay. The balance between the kinetic and
gravitational energies gives :

\begin{equation}  \label{ec}
\frac{N}{R} \sim \frac{G_4 M^2}{R} \, .
\end{equation}com
From here we get:

\begin{equation}
M \sim \frac{1}{G_4 m} \sim \frac{M_{Pl}^2}{m} \, .
\end{equation}
For example, if $m \sim 300$GeV then $M \sim 10^9$kg. These
states are known in literature as mini-soliton stars.

Consider now class I soliton model in the context of large extra
dimensions. Since the linear dimensions of this kind of
non-topological solitons  are smaller than the compactification
radius of the extra dimensions, this state is effectively a
spherically symmetric $(4+d)$-dimensional object. At these
distances, the gravitational strength is determined by a
fundamental scale $M_F$ rather than $M_{Pl}$. For simplicity, we
assume that the bulk cosmological constant is zero. The
equilibrium condition (\ref{ec}) is now

\begin{equation}
\frac{N}{R} \sim \frac{G_{4+d} M^2}{R^{1+d}} \, .
\end{equation}
From here we get:

\begin{equation}  \label{M}
M \sim \frac{R^d M_F^{d+2}}{m} \, .
\end{equation}
This formula is valid as long as $R$ is smaller than the
compactification radius of the extra dimensions, $r_c$. However,
there is an upper limit on the mass of such an object in a given
state. The critical mass $M_c$ for the formation of a black hole
can be estimated by substituting in (\ref{M}) an expression for
the Schwarzschild radius, $R_s$, of a $(4+d)$-dimensional black
hole \cite{bh}:

\begin{equation} \label{Rs}
R_s \sim \frac{1}{M_F} \left( \frac{M}{M_F}
\right)^{\frac{1}{d+1}} \, .
\end{equation}
We get:

\begin{equation}
M_c \sim \left( \frac{M_F}{m} \right)^{d+1} M_F \, .
\end{equation}

Note that, if we take $d=0$ and replace $M_F$ with $M_{Pl}$, we recover $%
(3+1)$-dimensional results. It is interesting that for the generic
values of parameters in the model, say $m \sim M_F \sim 1$TeV, the
critical mass is $M_c \sim 1$TeV and $R_s \sim 1$Tev$^{-1}$. However,
if the classical analysis of these states
is to be believed one need to choose parameters in the model which yield
the soliton mass much higher than the fundamental quantum gravity scale $M_F$.


Let us now examine $(3+1)$-dimensional class II soliton states. A
model which gives rise to this class of non-topological solitons
solutions contains a complex scalar field $\Phi$ whose action is
invariant under (\ref{u1}). Beside this, it has to satisfy an
additional condition --- that in the absence of the gravitational
field, the theory has non-topological soliton solutions. This can
be achieved by introducing a new Hermitian scalar field $\chi$ to
the Lagrangian (\ref{L}) (with $d=0$) with the potential:

\begin{equation}
V(\chi) = \frac{1}{2} \mu^2 \chi^2 \left( 1 - \frac{\chi}{\chi_0}
\right)^2 \, ,
\end{equation}
where $\mu$ is the mass of the field $\chi$. The normal vacuum state is $%
\chi =0$, while $\chi =\chi_0$ is the false vacuum state. Without
gravity, the theory contains a non-topological soliton solution.
The soliton contains an interior in which $\chi \sim \chi_0$, a
shell of width $\mu^{-1}$ over which $\chi$ changes from $\chi_0$
to $0$, and an exterior which is essentially the vacuum. The
field $\Phi$, with conserved particle number $N$ is confined to
the interior. The only role of the field $\chi$ is to limit the
space volume available to the filed $\Phi$. In opposite, the only
solutions for the filed $\Phi$ in the absence of gravity would be
the plain waves.

Consider a shell of radius $R$ with above mentioned
configuration. The kinetic energy inside the shell is:
\begin{equation}
E_k \sim \frac{N}{R} \, .
\end{equation}
Since the value of $\chi$ from interior to exterior changes from
$\chi_0$ to $0$, the shell contains a surface energy
\begin{equation} E_s \sim s R^2 \, ,
\end{equation}
where the surface tension $s \sim \mu \chi_0^2$. The equilibrium
radius can be found by minimizing the total energy $E = E_k +
E_s$ with respect to $R$. From there we find the minimum of $E$
which is the soliton mass, $M$. Assuming $\chi_0 \sim \mu$ this
is $ M \sim \mu^3 R^2$. Since the total conserved quantum number
is $N \sim s R^3$, the total soliton mass is $M \propto N^{2/3}$.
The exponent of $N$ is less than unity which implies, that for
large enough $N$, the total soliton mass is always less than the
sum of free particles masses. Therefore this soliton configuration is
stable.

Gravitational effects become important when $R$ approaches the
Schwarzschild radius. By setting $R_s \sim G_4 M$  we estimate
the critical mass, $M_c$, for the formation of a black hole:

\begin{equation} M_c \sim \left( \frac{M_{Pl}}{\mu} \right)^4 \mu
\end{equation}
For $\mu \sim 300$GeV we get $M_c \sim 10^{12} M_{Sun}$ and $R_s
\sim 1$ light month. This kind of solution is described in
\cite{ss}.

Consider now class II  soliton model in the context of large extra
dimensions. If the linear dimension, $R$, of a composite soliton
object  is much larger than the compactification radius of the
extra dimensions, then the order of magnitude estimate presented
above will not change much. However, on distances smaller than
the compactification radius, the gravitational interaction and
the functional dependence of the surface energy is quite
different than in $(3+1)$-dimensional case. This will lead to the
existence of the equilibrium configuration of the soliton states
whose liner dimensions are microscopic.

Consider again a shell of radius $R$ which is smaller than the
compactification radius of the extra dimensions. The kinetic
energy inside the shell remains $ E_{k}\sim N/R $. But, the
surface energy is $ E_{s}\sim sR^{2+d}$, where the surface
tension $s\sim \mu \chi _{0}^{2+d}$. By minimizing the total
energy $E=E_{k}+E_{s}$ with respect to $R$ and assuming $\chi
_{0}\sim \mu $ we get:

\begin{equation}  \label{Mssl}
M \sim \mu^{3+d} R^{2+d} \, .
\end{equation}

Note that $M \propto N^{\frac{2+d}{3+d}}$, and as in the case of
$(3+1)$-dimensions, the exponent of $N$ is less than unity which
implies the stability of the composite soliton state for large
enough $N$.

Gravitational effects become important when $R$ approaches the
$(4+d)$-dimensional Schwarzschild radius (\ref{Rs}). Thus, we
find the critical mass for the formation of a black hole:

\begin{equation}  \label{ssse}
M_c \sim \left( \frac{M_{F}}{\mu} \right)^{(d+2)^2} \mu \, .
\end{equation}
Note again that, if we take $d=0$ and replace $M_F$ with
$M_{Pl}$, we recover $(3+1)$-dimensional results. Because of the
form of the exponent in (\ref{ssse}), the critical mass is very
sensitive to the choice of the parameters in the model. Unlike in
$(3+1)$-dimensional models of class II solitons, there is an
equilibrium state of a composite soliton state which has
microscopic dimensions.  Like in case of class I soliton states,
it is possible to set parameters close to the generic ones ---
for example $\mu \sim M_{F}\sim 1$TeV --- which gives $M_{c}\sim 1$TeV and $R_s
\sim 1$TeV$^{-1}$. However, note again that if the classical analysis of
these states is to be believed one has to choose parameters which yield
the soliton mass much higher than the fundamental quantum gravity scale $M_F$.
Also, a large number $N$ of individual particles is needed  in
order to provide the classical stability of a composite soliton state which
will result in the large total soliton mass.

Although a careful numerical analysis is required in order to
estimate the total cross section production of these states in
near future accelerator experiments (like the Large Hadron
Collider with the center of mass energy of $14$TeV or the Very
Large Hadron Collider with the center of mass energy of
order $100$TeV ) it is conceivable that for some values of the
parameters in the model the cross section would not be negligible.


In conclusion, we considered non-linear scalar filed theory
models in the presence of gravity in the context of brane world
models with large compact extra dimensions. We showed that new
exotic non-topological soliton-like particles can exist.
One class of these states arises in simple models which
contain soliton solutions only in the presence of gravity.
The other class arises in a different type of models which contain
soliton solutions even in the absence of gravity. In this class, it is possible
to have equilibrium states of these solitons  which have both macro- and microscopic
dimensions. The latter was not possible in $(3+1)$-dimensional models.

Depending on the abundance of these new states of matter, they
can be important dark matter candidates not only because they are
cold and massive but also because of the fact that great part of
the dark matter must be of non-baryonic origin. It will be
interesting to study the role of these states in cosmology and
their contribution to entropy and energy density in our universe.

\bigskip

\vspace{12pt} {\bf Acknowledgments}:\ \  The author is grateful to
Valeri Frolov, Glenn Starkman, Tanmay Vachaspati and Bruce
Campbell for stimulating discussions. This work was partly
supported by the Natural Sciences and Engineering Research
Council of Canada. The author is also grateful to the Killam Trust
for its financial support.

\end{document}